\documentclass[twoside]{article}\usepackage[margin=1in,centering]{geometry}
\newcommand\aut {Leonid A.~Levin} \newcommand\ttl {Fundamentals of Computing}
\usepackage[utf8]{inputenc}\usepackage {amsmath,amssymb,amsthm,bm,mmap}
\usepackage[unicode,colorlinks]{hyperref}
\begin{document}\frenchspacing

 \newcommand\hreff [1]{{\footnotesize\url{https://#1}}}
 \newcommand\drop[1]{} \newcommand\trm[1]{{\bf\em #1}}
 \newcommand\Z{{\mathbb Z}}\newcommand\edf{{\mathrel{\mathop:}=\,}}

\author {\aut\ (\hreff {www.cs.bu.edu/fac/lnd/})}
 \title{\vspace*{-7pc}\ttl\vspace*{-10pt}}\date{}\maketitle

\vspace{-28pt}\begin{abstract} These are notes for the course CS-172 I first
taught in the Fall 1986 at UC Berkeley and subsequently at Boston University.
The goal was to introduce the undergraduates to basic concepts of Theory of
Computation and to provoke their interest in further study. Model-dependent
effects were systematically ignored. Concrete computational problems were
considered only as illustrations of general principles.

The notes are skeletal: they do have (terse) proofs, but exercises, references,
intuitive comments, examples are missing or inadequate. The notes can be used
for designing a course or by students who want to refresh the known material or
are bright and have access to an instructor for questions.\\ Each subsection
takes about a week of the course. Versions of these notes appeared in \cite
{SGN}. \end {abstract}

\paragraph {Acknowledgments.} {\small I am grateful to the University of
California at Berkeley, its MacKey Professorship fund and Manuel Blum who made
possible for me to teach this course. The opportunity to attend lectures of M.
Blum and Richard Karp and many ideas of my colleagues at BU and MIT were very
beneficial for my lectures. I am also grateful to the California Institute of
Technology for a semester with light teaching load in a stimulating environment
enabling me to rewrite the students' notes. NSF grants \#DCR-8304498,
DCR-8607492, CCR-9015276 also supported the work. And most of all I am grateful
to the students who not only have originally written these notes, but also
influenced the lectures a lot by providing very intelligent reactions and
criticism.

 \tableofcontents \vfill\noindent Copyright
 \textcircled{c} $\number\year$ by the author. \hfill Last revised: \today.}

\newpage\part {Basics} \label{basics}
 \section {Models of Computations;
 Polynomial Time \& Church's Thesis.}\label {models}

\subsection {Deterministic Computation.}

Sections~\ref {models},\ref{diagonal} study deterministic computations.
Non-deterministic aspects of computations (inputs, interaction, errors,
randomization, etc.) are crucial and challenging in advanced theory and
practice. Defining them as an extension of deterministic computations is
simple. The latter, however, while simpler conceptually, require elaborate
models for definition. These models may be sophisticated if we need a precise
measurement of all required resources. However, if we only need to define
what is computable and get a very rough magnitude of the needed resources,
all reasonable models turn out equivalent, even to the simplest ones.
We will pay significant attention to this surprising and important fact.
The simplest models are most useful for proving negative results
and the strongest ones for positive results.

We start with terminology common to all models, gradually making it more
specific to those we actually study. We represent \trm {computations} as
graphs: the edges reflect various relations between nodes (\trm {events}).
Nodes, edges have attributes: labels, states, colors, parameters, etc.
(affecting the computation or its analysis). \trm {Causal} edges run from each
event to all events essential for its occurrence or attributes. They form a
directed acyclic graph (though cycles may be added artificially to mark the
external input parts of the computation).

We will study only \trm {synchronous} computations. Their nodes have a \trm
{time} parameter. It reflects logical steps, not necessarily a precise value of
any physical clock. Causal edges only span short (typically, $\le 3$ moments)
time intervals. One event among the causes of a node is called its \trm
{parent}. \trm {Pointer} edges connect the parent of each event to all its
other possible causes and reflect connections that allow simultaneous events to
interact and have a joint effect. Pointers with the same source have different
labels. The (labeled) subgraph of events/edges at a given time is an instant
memory \trm {configuration} of the model.

Each non-terminal configuration has \trm {active} nodes/edges around which
it may change. The models with only a small active area at any step of
the computation are \trm {sequential}. Others are called \trm {parallel}.

\paragraph {Complexity.} We use the following measures of computing resources of
a machine $A$ on input $x$:

\trm {Time}: The greatest depth $D_{A(x)}$ of causal chains is the number
of computation steps. The volume $V_{A(x)}$ is the combined number of active
edges during all steps. Time $T_{A(x)}$ is used (depending on the context)
as either depth or volume, which are close for sequential models.
 Note that time complexity is robust only up to a constant factor:
a machine can be modified into a new one with a larger alphabet of labels,
representing several locations in one. It would produce identical results
in a fraction of time and space (provided that the time limits suffice
for transforming the input and output into the other alphabet).

\trm {Space}: $S_{A(x)}$ or $S_A(x)$ of a synchronous computation is the
greatest (over time) size of its configurations. Sometimes excluded are
nodes/edges unchanged since the input.

\paragraph {Growth Rates} (typically expressed as functions
of bit length $n=\|x,y\|$ of input/output $x$/$y$):

 $O,\Omega$: $f(n)=O(g(n))$\footnote
 {This is a customary but somewhat misleading notation.
 The clear notations would be like $f(n)\in O(g(n))$}
 $\iff g(n)=\Omega(f(n)) \iff \sup_n\frac{f(n)}{g(n)} <\infty$.

 $o,\omega: f(n)=o(g(n)) \iff g(n)=\omega(f(n))
 \iff \lim_{n \to \infty}\frac{f(n)}{g(n)}=0$.

$\Theta: f(n)=\Theta(g(n)) \iff$ ($f(n)=O(g(n))$ and $g(n)=O(f(n))$).

Here are a few examples of frequently appearing growth rates: negligible $(\log
n)^{O(1)}$; moderate $n^{\Theta(1)}$ (called polynomial or P, like in P-time);
infeasible: $2^{n^{\Omega(1)}}$, also $n!=(n/e)^n\sqrt{\pi(2n{+}1/3)+
\varepsilon/n}$, $\varepsilon{\in}[0,\,.1]$.\footnote
 {A rougher estimate follows by computing $\ln n!=(n{+}.5)\ln n-n+O(1)$ using
 that $|\sum_{i=2}^ng(i)-\int_{1.5}^{n+.5}g(t)\,{\bf d}t|$ is bounded by
 the total variation $v$ of $g'/8$. So $v<1/12$ for monotone $g'(t)=\ln'(t)=1/t$
 and the $O(1)$ is $1.5(1{-}\ln1.5)+\varepsilon$, $\varepsilon{\in}[0,\,.1]$.}

The reason for ruling out exponential (and neglecting logarithmic) rates
is that the visible Universe is too small to accommodate exponents.
Its radius is about 46.5 giga-light-years $\sim2^{204}$ Plank units.
 A system of $\gg R^{1.5}$ atoms packed in $R$ Plank Units radius
 collapses rapidly, be it Universe-sized or a neutron star.
 So the number of atoms is $<2^{306}\ll 4^{4^4}\ll 5!!$.

\newpage\subsection {Rigid Models.}

\trm {Rigid} computations have another node parameter: \trm {location} or \trm
{cell}. Combined with time, it designates the event uniquely. Locations have
\trm {structure} or \trm {proximity} edges between them. They (or their short
chains) indicate all neighbors of a node to which pointers may be directed.

\subsubsection* {Cellular Automata (CA).}

CA are a parallel rigid model. Its sequential restriction is the \trm {Turing
Machine (TM)}. The configuration of CA is a (possibly multi-dimensional)
grid with a finite, independent of the grid size, alphabet of \trm {states}
to label the events. The states include, among other values, pointers to
the grid neighbors. At each step of the computation, the state of each cell
can change as prescribed by a \trm {transition} function (also called program)
applied to the previous states of the cell and its pointed-to neighbors.
The initial state of the cells is the input for the CA.
All subsequent states are determined by the transition function.

An example of a possible application of CA is a VLSI (very large scale
integration) chip represented as a grid of cells connected by wires (chains
of cells) of different lengths. The propagation of signals along the wires
is simulated by changing the state of the wire cells step by step. The clock
interval can be set to the time the signals propagate through the longest wire.
This way the delays affect the simulation implicitly.

\paragraph {An example: the Game of Life (GL).} GL is a plane grid of cells,
each holds a 1-bit state (dead/alive) and pointers to the 8 adjacent cells.
A cell remains dead or alive if the number $i$ of its live neighbors is $2$.\\
 It becomes (or stays) alive if $i{=}3$.
 In all other cases it dies (of overcrowding or loneliness) or stays dead.

A \trm {simulation} of a machine $M_1$ by $M_2$ is a correspondence
between memory configurations of $M_1$ and $M_2$ which is preserved during
the computation (may be with some time dilation). Such constructions show that
the computation of $M_1$ on any input $x$ can be performed by $M_2$ as well.
GL can simulate any CA (see a sketch of an ingenious proof
in the last section of \cite{ww}) in this formal sense:

We fix space and time periods $a,b$. Cells $(i,j)$ of GL are mapped to cell
$(\lfloor i/a\rfloor,\lfloor j/a\rfloor)$ of CA $M$ (compressing $a\times a$
blocks). We represent cell states of $M$ by states of $a\times a$ blocks
of $GL$. This correspondence is preserved after any number $t$ steps
of $M$ and $b t$ steps of $GL$ regardless of the starting configuration.

\subsubsection* {Turing Machines (TMs).}

TM is a {\em minimal CA}. Its configuration -- \trm{tape} -- is an infinite
to the right chain of cells.\\ Each state of a cell has a pointer to one of
the cell's two adjacent neighbors. No adjacent cells can both\\ point away from
each other. Only the two cells pointing at each other are \trm {active},
i.e. can change state.\\ The cell that just turned its pointer is the TM's
moving head working on the tape symbol - its target.\\ The input is an array
of non-blank cells (only one is rightward) followed by blanks at the right.

Another type of CA represents a TM $A$ with several non-communicating heads. At
most $O(1)$ heads fit in a cell. They can vanish, split, or merge only in the
first cell (which, thus, controls the number of active cells). The input $x$
makes an unchangeable ``ink" part of each cell's state. The rest of the cell's
state is in ``pencil" and can be changed by $A$. The computation halts when
all heads drop off. The output $A(x)$ is the pencil part of the tape's state.
This model has convenient theoretical features. E.g. with linear (in $T$)
number ($\|p\|^2T$) of state changes (volume) one can solve the \trm {Bounded
Halting Problem} $H(p,x,T)$: find out whether the machine with a program $p$
stops on an input $x$ within volume $T$ of computation (see \ref{compress}).

\vspace{-8pt}\paragraph {Exercise:} Find a method to transform any given
multi-head TM $A$ into another one $B$ such that the value of the output of
$B(x)$ (as a binary integer) and the volumes of computation of $A(x)$ and of
$B(x)$ are all equal within a constant factor (for all inputs $x$). {\bf Hint:}
$B$-cells may have a field to simulate $A$ and maintain (in other fields) two
binary counters $h$ (with $\Theta(1)$ density of heads) for the number of heads
of $A$ and $v$ for $A$'s volume. Their least significant digits are at the
leftmost cell. $h$ adds its most significant digit to the same position in $v$
at each step of $A$. To move the carry $1$ on $v$ a head is borrowed from $h$.
These $1$-heads move right in $v$ till they empty their carry into its $0$
digit. Then empty $0$-heads move back to $h$ in a separate field/track,
possibly first continuing right to find a free slot in this return track.
 (The heads area in $v$ extends to $k$-th cell only by dropping the
 carry there, with frequency $O(2^{-k})$. Then it shrinks to $O(1)$ in
 $O(k)$ steps since heads enter it slower than they move away.)
 Borrowed or returned heads make low or high head-density areas
 in $h$ which shift left until absorbed at the leftmost cell.

\newpage\subsection {Pointer Machines.}

The memory configuration of a \trm {Pointer Machine (PM)}, called \trm {pointer
  graph}, is a finite directed labeled multigraph. One node $R$ is marked as
\trm {root} and has directed paths to all nodes. Nodes can \trm {see} and
change the configuration of their out-neighborhood of constant (2 suffices)
depth. Edges (\trm {pointers}) are labeled with \trm {colors} from a finite
alphabet common to all graphs handled by a given program. The pointers coming
out of a node must have different colors (which bounds the outdegree). Some
colors are designated as \trm {working} and not used in inputs/outputs. One of
them is called \trm {active}, as also are pointers carrying it and nodes seeing
them. Active pointers must have inverses, form a tree to the root, and can be
dropped only in leaves.

All active nodes each step execute an identical \trm {program}.
At its first \trm {pulling} stage, node $A$ acquires copies of all pointers
of its children using ``composite" colors: e.g., for a two-pointer path
$(A,B,C)$ colored $x,y$, the new pointer $(A,C)$ is colored $xy$,
or an existing $z$-colored pointer $(A,C)$ is recolored $\{z,xy\}$.
$A$ also spawns a new node with pointers to and from it. Next, $A$ transforms
the colors of its set of pointers, drops the pointers left with composite
colors, and vanishes if no pointers are left. Nodes with no path from the root
are forever invisible and considered dropped. The computation is initiated by
inserting an active loop-edge into the root. When no active pointers remain,
the graph, with all working colors dropped, is the output.

{\bf Exercise:} Design a PM transforming the input graph into the same one with
two extra pointers from each node: to its parent in a BFS spanning tree and to
the root. Hint: Nodes with no path {\em to} the root can never be activated.
but can be copied with pointers, copies connected to the root, the original
input removed.

PM can be \trm {parallel, PPM} \cite{ba} or \trm {sequential, SPM}.
SPM differ in that only pointers to the root, their sources, and nodes
that have pointers with inverses to these sources can be active.

A \trm {Kolmogorov} or \trm {Kolmogorov-Uspenskii} Machine (KM) \cite{ku},
is a special case of Pointer Machine \cite{pm} with the restriction
that all pointers have inverses. This implies the bounded in/out-degree
of the graph which we further assume to be constant.

\trm {Fixed Connection} Machine (FCM) is a variant of the PKM with the
restriction that pointers once created {\em cannot} be removed, only
re-colored. So when the memory limits are reached, the pointer structure
freezes, and the computation can be continued only by changing the colors of
the pointers.

PPM is the most powerful model we consider: it can simulate the others
in the same space/time. E.g., cellular automata make a simple special
case of a PPM which restricts the Pointer Graph to be a grid.

\paragraph {Exercise.} Design a machine of each model (TM, CA, KM, PPM)
which determines if an input string $x$ has a form $ww$, $w\in\{a,b\}^*$.
Analyze time (depth) and space. KM/PPM takes input $x$ in the form of
colors of edges in a chain of nodes, with root linked to both ends.
The PPM nodes also have pointers to the root. Below are hints for TM,SPM,CA.
The space is $O(\|x\|)$ in all three cases.

{\bf Turing and Pointer Machines.} TM first finds the middle of $ww$
by capitalizing the letters at both ends one by one. Then it compares
letter by letter the two halves, lowering their case. The complexity is:
$T(x)=O(\|x\|^2)$. SPM acts similarly, except that the root keeps and updates
the pointers to the borders between the upper and lower case substrings.
This allows constant time access to these borders. So, $T(x){=}O(\|x\|)$.

{\bf Cellular Automata.} The computation starts with the leftmost cell
sending right two signals. Reaching the end the first signal turns back.
The second signal propagates three times slower, so they meet in the middle of
$ww$ and disappear. While alive, the second signal copies the input field $i$
of each cell into a special field $c$. The $c$ symbols will try to move right
whenever the next cell's $c$ field is blank. So the chain of these symbols
alternating with blanks will start moving right from the middle of $ww$.
Upon reaching the end they will push the blanks out and pack themselves
back into a copy of the left half of $ww$ shifted right. When a $c$ symbol
does not have a blank at the right to move to, it compares itself with
the $i$ field of the same cell. If they differ, a signal is generated which
halts all activity and rejects $x$. If all comparisons are successful,
the last $c$ generates the accepting signal. The depth is: $T(x)=O(\|x\|)$.

\newpage \subsection {Simulation.}

We have considered several models of computation. We will see now how
the simplest of them -- Turing Machine -- can simulate all others:
these powerful machines can compute no more functions than TM.

\paragraph {Church-Turing Thesis} is a generalization of this conclusion:
TMs can compute every function computable in any thinkable physical model
of computation. This is not a math theorem because the notion of model
is not formally specified. But the long history of studying ways
to design real and ideal computing devices makes it very convincing.
Moreover, this Thesis has a stronger \trm {Polynomial Time} version
which bounds the volume of computation required by that TM simulation
by a polynomial of the volume used by the other models. Both forms of
the Thesis play a significant role in foundations of Computer Science.

\paragraph {PKM Simulation of PPM.} For convenience, we assume all PPM nodes
have pointers to root. PPM configuration is represented in PKM with extra
colors $l,r,u$ used in a $u$-colored binary tree added to each node $X$ so that
all (unlimited in number) PPM pointers to $X$ are reconnected to its leaves,
and inverses, colored $l,r$, added to all pointers. The number of pointers
increases at most 4 times. To simulate PPM, $X$ gets a binary \trm {name}
formed by the $l,r$ colors on its path through the root tree, and broadcasts it
down its own tree. For pulling stage $X$ extends its tree to double depth and
merges (with combined colors) its own pointers to nodes with identical names.
Then $X$ re-colors its pointers as PPM program requires and rebalances its
tree. This simulation of a PPM step takes polylogarithmic parallel time.

\paragraph {TM Simulation of PKM.} We assume the PKM keeps a constant degree
and a roughly balanced root tree (to yield short node names as described
above). TM tape reflects its configuration as the list of all pointers sorted
by source name, then by color. The TM's transition table reflects the PKM
program. To simulate PKM's pulling stage TM creates a copy of each pointer and
sorts copies by their sinks. Now each pointer, located at source, has its copy
near its sink. So both components of 2-pointer paths are nearby: the special
double-colored pointers can be created and moved to their sources by resorting
on the source names. The re-coloring stage is straightforward, as all relevant
pointers having the same source are located together. Once the root has no
active pointers, the Turing machine stops and its tape represents the PKM
output. If a PPM computes a function $f(x)$ in $t(x)$ steps, using $s(x)$
nodes, the simulating TM uses space $S = O(s\log s)$, ($O(\log s)$ bits for
each of $O(s)$ pointers) and time $T= O(S^2)t$, as TM sorting takes quadratic
time.

\paragraph {Squaring matters !} {\small TM cannot outperform Bubble Sort.
Is its quadratic overhead a big deal? In a short time all silicon gates
on your PC run, say, $X{=} 10^{23}{\sim}2^{2^{6.25}}$ clock cycles combined.
Silicon parameters double almost annually. Decades may bring micron-thin things
that can sail sunlight in space in clouds of great computing and physical
(light beam) power. Centuries may turn them into a Dyson Sphere enveloping
the solar system. Still, the power of such an ultimate computer is limited
by the number of photons the Sun emits per second: $Y{\sim} 2^{2^{7.25}}{=}X^2$.
Giga-years may turn much of the known universe into a computer,
but its might is still limited by its total entropy $2^{2^{8.25}}{=}Y^2$.}

\paragraph {Faster PPM Simulations.} Parallel Bubble-Sort on CA or Merge-Sort
on sequential FCM take nearly linear time. Parallel FCM can do much better
\cite{ofm}. It represents and updates pointer graphs as the above TM.
All steps are straightforward to do locally in parallel polylog time except
sorting of pointers. We need to create a fixed connection sorting network.
Sophisticated networks sort arbitrary arrays of $n$ integers in $O(\log n)$
parallel steps. We need only a simpler polylog method. Merge-Sort splits
an array of two or more entries in two halves and sorts each recursively.
Batcher-Merge combines two sorted lists in $O(\log n)$ steps.

{\bf Batcher Merge.} A \trm {bitonic cycle} is the combination of two sorted
arrays (one may be shorter),\\ connected by max-to-max and min-to-min entries.
Entries in a contiguous half (\trm {high-half}) of the cycle\\
are $\ge$ than all entries in the other (\trm {low}) half.
Each half (with its ends connected) forms a bitonic cycle itself.

A \trm {flip} of an array entry is one with the highest address bit flipped.
Its \trm {shift} has the highest bit of its address cycle-shifted to the end.
Linking nodes in a $2^k$-nodes array to their flips and shifts forms a \trm
{Shuffle Exchange} graph. We merge-sort two sorted arrays given as a bitonic
cycle on such a graph as follows.

Comparing each entry with its flip (half-a-cycle away), and switching if
wrongly ordered, fits the high and low halves into respectively the first and
last halves of the array by shifting the dislocated segment of each (thus
rotating each cycle). This repeats for each half recursively (decrementing $k$
via graph's shift edges).

\newpage\section {Universal Algorithm; Diagonal Results.}
 \label{diagonal}

\subsection {Universal Turing Machine.}\label{ikeno}

The first computers were hardware-programmable. To change the function
computed, one had to reconnect the wires or even build a new computer. John von
Neumann suggested using Turing's Universal Algorithm. The function computed can
be then specified by just giving its description (program) as part of the input
rather than by changing the hardware. This was a radical idea, since in the
classical mathematics universal functions do not exist (as we will see in
Sec.~\ref{gdl}).

Let $R$ be the class of all TM-computable functions: total (defined for all
inputs) and partial (which may diverge). Surprisingly, there is a universal
function $u$ in $R$. For any Turing Machine $M$ that computes $f{\in}R$ in time
$T$ and space $S$, $u$ uses a program $m$ of length $c$ listing the commands
and initial head state of $M$. Then $u(m x)$ simulates $M(x)$ in time $c^2 T$
and space $S{+}c$. It operates in cycles, each simulating one step of $M(x)$.
After $i$ steps of $M(x)$, let $s_i$ be the head's state, $l_i$ be the left
from it part of the tape; $r_i$ be the rest of the tape. After $i$ cycles
$u(m x)$ has the tape configuration $t_i=l_i m s_i r_i$ and looks up $m$
to find the command corresponding to the state $s_i$ and the first symbol
of $r_i$. It modifies $t_i$ accordingly. When $M(x)$ halts, $u(m x)$ erases
the (penciled) $ms_i$ from the tape and halts too. Universal Multi-head TM
works similarly but can also determine in time $O(t(x))$ whether
it halts in $t$ steps (given $x,t(x)$ and an appropriate program).

{\bf Exercise.} Design a universal multi-head TM with a constant factor
overhead.\\ Hint: When heads split or merge in the first cell,
the room $u$ needs for their programs\\
creates sparse or dense content regions that propagate right (sparse faster).

We now describe in detail a simpler but slower universal \cite{Ikeno} TM $U$.
It simulates any other TM $M$ that uses only $0,1$ bits on the tape. (Other
alphabets are encoded as bit strings for this.) $M$ lacks the blank symbol that
usually marks the end of input. Thus input needs to be given in some prefixless
form, e.g. with a padding that encodes input length $l=\|x\|$ as a binary
string preceded by a string of $2\|l\|$ zeros. In the cells carrying this
padding, two counters are initiated that monitor the distance to both ends of
the used part of $M$'s tape (initially the input). $M$'s head moving on the
tape pulls these counters along and keeps them updated. When the right end of
the used tape is reached, any subsequent characters are treated as blanks.

\noindent\parbox {24pc} {\hspace*{1pc} $U$ has 11 states + 6 symbols; its
transition table is at the right. It shows the states and tape digits only when
changed, except that the prime is always shown. The head is on the tape: lower
case states look left, upper case -- right. The external choice, halt, etc.
commands are special states for $M$; for $U$ they are shown as A/B or =. $U$
works in cycles, simulating one transition of $M$ each. The tape is infinite to
the right (the leftward head in the leftmost cell halts). It consist of $0/1$
segments, each preceded with a *. Some symbols are primed. The rightmost
infinite part of one or two segments is always a copy of $M$'s tape. The other
segments describe one transition each: a command $(s,b)\to (s',b',d)$ for $M$
to change state $s$ to $s'$, tape bit $b$ to $b'$ and turn left or right.}
 \hfill${\begin{tabular} {|c||c|c|c|c|c|c|}\hline
    &1 &0 &* &1'&0'& *'\\ \hline\hline
  A &f &f &e0&  &  &   \\ \hline
  B &F &F &e1&  &  &   \\ \hline
 f,F&b*&a*& F&  &  & c \\ \hline
  D &d'&--&  &  &  & e'\\ \hline
  E &= &--&' &' &' & e'\\ \hline
  d &' &' &  &' &' & D \\ \hline
  b &' &' &' &  &a'& D \\ \hline
  a &' &' &' &b'&F & E'\\ \hline
  c &' &' &  &= &F & E'\\ \hline
  e &' &' &' &B &A &A/B\\ \hline\end{tabular}}$

The transition segments are sorted in order of $(s,b)$ and never change, except
for priming. Each transition is represented as $*S d b$, where $b$ is the bit to
write, $d$ the direction $R{=}0/L{=}1$ to turn. $S$ points to the next state
represented as $1^k$, if it is $k$ segments to the left, or $0^k$ (if to the
right). Each cycle starts with $U$'s head in state $F$ or $f$, located at the
site of $M$'s head. Primed are the digits of $S$ in the prior command and all
digits to their left. An example of the configuration:
 \mbox{$ *'0'0'0'1'0'*'0'0'0'0'01*011*\ldots *00\ \fbox{head}\ 00\ldots$}

$U$ first reads the bit of an $M$'s cell changing the state from $F$ or $f$
to $a/b$, puts a * there, moves left to the primed state segment $S$, finds
from it the command segment and moves there. It does this by repeatedly priming
nearest unprimed * and 1s of $S$ (or unpriming 0s) while alternating the states
$c/F$ or $d/D$. When $S$ is exhausted, the target segment $\|S\|+b$ stars away
is reached. Then $U$ reads (changing state from $e$ to $A$ or $B$) the
rightmost symbol $b'$ of the command, copies it at the * in the $M$ area,
goes back, reads the next symbol $d$, returns to the just overwritten
(and first unprimed) cell of $M$ area and turns left or right. As CA, $M$
and $U$ have in each cell three standard bits: present and previous pointer
directions and a ``content" bit to store M's symbol. In addition $U$ needs
just one ``trit" of its own!\footnote {See its simulator at \hreff
   {www.cs.bu.edu/teaching/cs332/TM/simulator/}}

\newpage \subsection {Uncomputability; Goedel Theorem.}\label{gdl}

\subsubsection* {Universal and Complete Functions.}\vspace{-2pt}

Notations: Let us choose a special mark and after its $k$-th occurrence, break
any string $x$ into Prefix$_k(x)$ and Suffix$_k(x)$. Let $f^+(x)$ be
$f($Prefix$_k(x)\ x)$ and $f^-(x)$ be $f($Suffix$_k(x))$. We say $u$ $k$-\trm
{simulates} $f$ iff for some $p=$Prefix$_k(p)$ and all $s$, $u(ps)=f(s)$. The
prefix can be intuitively viewed as a program which simulating function $u$
applies to the suffix (input). We also consider a symmetric variant of relation
``$k$-simulate" which makes some proofs easier. Namely, $u$ $k$-\trm
{intersects} $f$ iff $u(ps)=f(ps)$ for some prefix$_k$ $p$ and all $s$. E.g.,
length preserving functions can intersect but cannot simulate one another.

We call \trm {universal} for a class $F$, any $u$ which $k$-simulates
all functions in $F$ for a fixed $k$. When $F$ contains $f^-,f^+$ for each
$f\in F$, universality is equivalent to (or implies, if only $f^+\in F$) \trm
{completeness}: $u$ $k$-intersects all $f\in F$. Indeed, $u$ $k$-simulates $f$
iff it $k$-intersects $f^-$; $u$ $2k$-intersects $f$ if it $k$-simulates $f^+$.

{\bf Exercise:} Describe {\bf explicitly} a function, {\bf complete}
for the class of all {\em linear} (e.g., $5x$ or $23x$) functions.

A \trm {negation} of a (partial or total) function $f$ is the total predicate
$\neg f$ which yields $1$ iff $f(x){=}0$ and yields $0$ otherwise. Obviously,
no closed under negation class of functions contains a complete one. So, there
is no universal function in the class of all (computable or not) predicates.
This is the well known Cantor Theorem that the set of all sets of strings (as
well as the sets of all functions, reals etc.) is not countable.

\subsubsection* {Goedel's Theorem.}\vspace{-1pt}

There is no complete function among the {\bf total} computable ones,
as this class is closed under negation.\\ So the universal in $R$ function $u$
(and $u_2=(u\bmod2)$) has no total computable extensions.

Formal proof systems are computable functions $A(P)$ which check if $P$ is an
acceptable proof and output the proven statement. $\vdash s$ means $s=A(P)$ for
some $P$. $A$ is \trm {rich} iff it allows computable translations $s_x$ of
statements ``$u_2(x)=0$", provable whenever true, and refutable ($\vdash\neg
s_x$), whenever $u_2(x)=1$. $A$ is \trm {consistent} iff {\bf at most} one of
any such pair $s_x,\neg s_x$ is provable, and \trm{complete} iff {\bf at least}
one of them always (even when $u(x)$ diverges) is. Rich consistent and complete
formal systems cannot exist, since they would provide an obvious total
extension $u_A$ of $u_2$ (by exhaustive search for $P$ to prove or refute
$s_x$). This is the famous Goedel's Theorem -- one of the shocking surprises of
the 20th century science. (Here $A$ is any extension of the formal Peano
Arithmetic; we skip the details of its formalization and proof of
richness.)\footnote
 {A closer look at this proof reveals another famous Goedel theorem:
 Consistency $C$ of $A$ (expressible in $A$ as divergence of the search
 for contradictions) is itself an example of unprovable $\neg s_x$. Indeed,
 $u_2$ intersects $1{-}u_A$ for some prefix $a$. $C$ implies that $u_A$ extends
 $u_2$ and, thus, $u_2(a),u_A(a)$ both diverge. So, $C{\Rightarrow\neg} s_a$.
 This proof can be formalized in Peano Arithmetic, thus $\vdash C\Rightarrow\
 \vdash\neg s_a$. But $\vdash\neg s_a$ implies $u_A(a)$ converges, so $\vdash C$
 contradicts $C$: Consistency of $A$ is provable in $A$ if and only if false !}

\paragraph {Recursive Functions.\label{rec}} Another byproduct is that the
Halting (of $u(x)$) Problem would yield a total extension of $u$ and, thus, is
not computable. This is the source of many other uncomputability results.
Another source is an elegant \trm {Fixed Point} Theorem by S. Kleene: any total
computable transformation $A$ of programs (prefixes) maps some program into an
equivalent one. Indeed, the complete/universal $u(ps)$ intersects computable
$u(A(p)s)$. This implies, e.g., Rice theorem: the only
computable invariant (i.e. the same on programs computing the same functions)
property of programs is constant ({\bf exercise}).

Computable (partial and total) functions are also called \trm {recursive} (due
to an alternative definition). Their ranges (and, equivalently, domains) are
called (recursively) \trm {enumerable} or \trm {r.e.} sets. An r.e. set with an
r.e. complement is called recursive (as is its yes/no characteristic function)
or \trm {decidable}. A function is recursive iff its graph is r.e. An r.e.
graph of a total function is recursive. Each infinite r.e. set is the range of
an injective total recursive function (``enumerating" it, hence the name r.e.).

We can reduce membership problem of a set $A$ to the one of a set $B$
by finding a recursive function $f$ s.t. $x\in A \iff f(x)\in B$.
Then $A$ is called \trm {m-} (or \trm {many-to-1-}) \trm {reducible} to $B$.
A more complex \trm {Turing} reduction is given by an algorithm which, starting
from input $x$, interacts with $B$ by generating strings $s$ and receiving
answers to $s\in?B$ questions. Eventually it stops and tells if $x\in A$.
R.e. sets (like Halting Problem) to which all r.e. sets can be m-reduced are
called r.e.-complete. One can show a set r.e.-complete (and, thus, undecidable)
by reducing the Halting Problem to it. So Ju.Matijasevich proved
r.e.-completeness of Diophantine Equations Problem: given a multivariate
polynomial of degree 4 with integer coefficients, find if it has integer roots.
 The above (and related) concepts and facts are broadly used in Theory of
Algorithms and should be learned from any standard text, e.g., \cite{rg}.

\newpage\subsection
{Intractability; Compression and Speed-up Theorems.}\label{compress}

The \trm {$t$-restriction} $u_t$ of $u$ aborts and outputs $1$ if $u(x)$ does
not halt within $t(x)$ steps, i.e. $u_t$ computes the \trm {$t$-Bounded Halting
Problem ($t$-BHP)}. It remains complete for the closed under negation class of
functions computable in $o(t(x))$ steps. ($O(\|p\|^2)$ overhead is absorbed by
$o(1)$ and padding $p$.) So, $u_t$ is not~in the class, i.e. cannot be computed
in time $o(t(x))$ \cite{ts}. (And neither can be any function agreeing~with
$t$-BHP on a \trm {dense} (i.e. having strings with each prefix) subset.)
E.g. $2^{\|x\|}$-BHP requires exponential time.

However for some trivial input programs the BHT can obviously be answered by a
fast algorithm. The following theorem provides another function $P_f(x)$ (which
can be made a predicate) for which there is only a finite number of such
trivial inputs. We state the theorem for the volume of computation of
Multi-Head Turing Machine. It can be reformulated in terms of time of Pointer
Machine and space (or, with smaller accuracy, time) of regular Turing Machine.

\paragraph {Definition:} A function $f(x)$ is \trm {constructible}
if it can be computed in volume $V(x)=O(f(x))$.

Here are two examples: $2^{\|x\|}$ is constructible, as $V(x)=O(\|x\|\log\|x\|)
\ll 2^{\|x\|}$.\\ Yet, $2^{\|x\|} + h(x)$, where $h(x)$ is $0$ or $1$,
depending on whether $U(x)$ halts within $3^{\|x\|}$ steps, is not.

\paragraph {Compression Theorem \cite{rb}.} For any constructible function $f$,
there exists a function $P_f$ such that for all functions $t$,
the following two statements are equivalent:
 \begin{enumerate}\item There exists an algorithm $A$ such that
 $A(x)$ computes $P_f(x)$ in volume $t(x)$ for all inputs $x$.
\item $t$ is constructible {\em and} $f(x)=O(t(x))$.\end{enumerate}

\paragraph {Proof.} Let \trm {$t$-bounded Kolmogorov Complexity} $K_t(i|x)$
of $i$ given $x$ be the length of the shortest program $p$ for the Universal
Multi-Head Turing Machine transforming $x$ into $i$ with $<t$ volume of
computation. Let $P_f(x)$ be the smallest $i$, with $2K_t(i|x)>\log(f(x)|t)$
for all $t$. $P_f$ is computed in volume $f$ by generating all $i$ of
low complexity, sorting them and taking the first missing. It satisfies
the Theorem, since computing $i{=}P_f(x)$ faster would violate the complexity
bound defining it. (Some extra efforts can make $P$ Boolean.)~$\qed$

\paragraph {Speed-up Theorem \cite{spd}.} There exists a total computable
predicate $P$ such that for any algorithm computing $P(x)$ in volume $t(x)$,
there exists another algorithm doing it in volume $O(\log t(x))$.

\vspace{6pt}\noindent Though stated here for exponential speed-up, this theorem
remains true with $\log$ replaced by any computable unbounded monotone function.
In other words, there is no even nearly optimal algorithm to compute $P$.

\paragraph {The general case.} So, the complexity of some predicates $P$ cannot
be characterized by a single constructible function $f$, as in Compression
Theorem. However, the Compression Theorem remains true (with harder proof) if
the requirement that $f$ is constructible is dropped (replaced with being
computable).\footnote
 {The proof stands if constructibility of $f$ is weakened to being
 \trm {semi-constructible}, i.e. one with an algorithm $A(n,x)$
 running in volume $O(n)$ and such that $A(n,x){=}f(x)$ if $n{>}f(x)$.
 The sets of programs $t$ whose volumes (where finite) satisfy
 either (1) or (2) of the Theorem (for computable $P,f$) are in $\Sigma^0_2$
 (i.e. defined with 2 quantifiers). Both generate monotone classes of
 constructible functions closed under $\min(t_1,t_2)/2$. Then any such class
 is shown to be the $\Omega(f)$ for some {\bf semi-constructible}~$f$.}\\
 In this form it is general enough so that every computable predicate
 (or function) $P$ satisfies the statement of the theorem with an appropriate
 computable function $f$. There is no contradiction with Blum's Speed-up,
 since the complexity $f$ (not constructible itself) cannot be reached.
 See a review in \cite{sm}.

\newpage \section {Games; Alternation; Exhaustive Search;
 Time vs. Space.} \label{games}

\subsection {How to Win.} \label{win}

In this section we consider a more interesting {\em provably} intractable
problem: playing games with full information, two players and zero sum. We will
see that even for some simple games there cannot be a much more efficient
algorithm than exhaustive search through all possible configurations.

The rules of an \trm {$n$-player game} $G$ are set by families $f,v$ of \trm
{information} and \trm {value} functions and a \trm {transition rule} $r$.
Each player $i\in I$ at each step participates in transforming a configuration
(game position) $x\in C$ into the new configuration $r(x,m),\ m:I\to M$ by
choosing a move $m_i= m(i)$ based only on his knowledge $f_i(x)$ of $x$.
The game proceeds until a \trm {terminal} configurations $t\in T\subset C$
is reached. Then $v_i(t)$ is the loss or gain of the $i$-th player.
Our games will have \trm{zero sum} $\sum v_i(t)=0$ and \trm {full information:}
$f_i(x)=x$, $r(x,m)=r'(x,m_{a(x)})$, where $a(x)$ points to
the \trm {active} player. We consider binary, two-players, no-draw games,
taking $0{\notin}C{\subset}\mathbb Z$, $M{\subset}\mathbb Z$, $T=I=\{\pm1\}$,
$a(x)={\bf sign}(x)$, $v_i(t)=a(t)i$, and $|r(x,m)|<|x|$.\footnote
 {Our examples will assure ``$<|x|$" by implicitly prepending
 non-terminal configurations with a counter of remaining steps.}

An example of such games is chess. Examples of games without full information
are card games, where only a part $f_i(x)$ (player's own hand) of the position
$x$ is known. Each player may have a \trm {strategy} providing a move for each
position. A strategy $S$ is \trm {winning} if it guarantees victory whatever
the opponent does, even if he knows $S$. We can extend $v_1$ on $T$ to $V$ on
all positions with a winning strategy for one side so that
$a(x)V(x)=\sup_m\{a(x)V(r(x,m))\}$. ($\sup\{\}$ taken as $-1$.)

\trm {Evaluating} or \trm {solving} a game, means computing $V$.
This ability is close to the ability to find a good move in a modified game.
Indeed, modify a game $G$ into $G'$ by adding a preliminary stage to it.\\
At this stage the player $A$ offers a starting position for $G$ and her
opponent $B$ chooses which side to play. Then $A$ may either start playing $G$
or decrement the counter of unused positions and offer another one.
Obviously, $B$ wins if he can determine the winning side of every position. If
he cannot while $A$ can, $A$ wins. Also, any game can be modified into one with
two moves: $M\subset \{0,1\}$ by breaking a string move into several bit-moves.
(A position of the new game consists of a position $x$ of the old one
and a prefix $y$ of a move. The active player keeps adding bits to $y$
until $m$ is complete and the next position generated by $r(x,m)$.)
 Evaluating such games is obviously sufficient for choosing the right move.

\paragraph {Theorem.} {\em Each position of any full information game
  has a winning strategy for one side.}

(This theorem \cite{NM} fails for games with \trm {partial} information:
either player\\ may lose if his strategy is known to the adversary. E.g.:
1. Blackjack (21); 2. Each player picks a bit;\\ their equality determines the
winner.) The game can be solved by playing all strategies against each other.
There are $2^n$ positions of length $n$, $(2^n)^{2^n} = 2^{n\times 2^n}$
strategies and $2^{n\times 2^{n+1}}$ pairs of them. For a 5-bit game that is
$2^{320}$. The proof of this Theorem gives a much faster
(but still exponential time!) strategy.

\paragraph {Proof.} Make the graph of all ${\le}\|x\|$ -bit positions and moves;
Set $V=0$; reset $V=v$ on $T$.\\ Repeat until idle: If $V(x)=0$, set
$V(x)=a(x)\sup_m\{a(x)V(r(x,m))\}$.\\ The procedure stops with empty
$V^{-1}(0)$ since $|r(x,m)|<|x|$ in our games keep decreasing. $\qed$

Games may be categorized by the difficulty to compute $r$. We will consider
only $r$ computable in\\ linear space $O(\|x\|)$. Then, the $2^{2\|x\|}$
possible moves can be computed in exponential time, say $2^{3\|x\|}$.\\
The algorithm tries each move in each step. Thus, its total running time
is $2^{3\|x\|+1}$: {\em extremely} slow\\ ($2^{313}$ for a 13-byte game) but
still {\em much} faster than the previous (double exponential) algorithm.

\paragraph {Exercise: the Match Game.} Consider 3 boxes with 3 matches each:
\mbox{{\Large \fbox{ ! ! ! } \fbox{ ! ! ! } \fbox{ ! ! ! }}}.\\
 The players alternate turns taking any {\em positive} number of matches from
 a {\em single} box. One cannot leave the table empty. Use the above algorithm
 to evaluate all positions and list the evaluations after each its cycle.

\paragraph {Exercise:} Modify the chess game by giving one side the right to
make (if it chooses to) an extra\\ move out of turn during the first 10 moves.
Prove that this side have a non-loosing strategy.

\newpage\subsection {Exponentially Hard Games.} \label{exp-g}

A simple example of a full information game is \trm {Linear Chess}, played on
a finite linear board. Each piece has a 1-byte type, including {\em loyalty}
to one of two sides: W (weak) or S (shy), {\em gender} M/F and a 6-bit {\em
 rank}. All cells of the board are filled and all W's are always on the left
of all S's. Changes occur only at the \trm {active} border where W and S meet
(and fight). The winner of a fight is determined by the following Gender Rules:

1. If S and W are of the same sex, W (being weaker) loses.

2. If S and W are of different sexes, S gets confused and loses.

The party of a winning piece A replaces the loser's piece B by its own piece C.
The choice of C is restricted by the table of rules listing all allowed
triples (ABC). We will see that this game {\em cannot} be solved in
a {\em subexponential} time. We first prove that (see \cite{CKS}) for
an artificial game. Then we reduce this \trm {Halting Game} to Linear Chess
showing that any fast algorithm to solve Linear Chess, could be used to solve
Halting Game, thus requiring exponential time. For Exp-Time Completeness
of regular (but $n\times n$) Chess, Go, Checkers see: \cite{chess,go}.

\subsubsection* {Exptime Complete Halting Game.} \label{halt-gm}

We use a universal Turing Machine $u$ (defined as 1-pointer cellular automata)
which halts only by its head rolling off of the tape's left end, leaving
a blank. Bounded Halting Problem BHP$(x)$ determines if $u(x)$ stops
(i.e. the leftmost tape cell points left) within $2^{\|x\|}$ steps. This cannot
be done in $o(2^{\|x\|})$ steps.\\ We now convert $u$ into the Halting Game.

\noindent\parbox{270pt} {The players are: $L$ claiming $u(x)$ halts in time
(and should have winning strategy iff this is true); His opponent is $S$.
The \trm {board} has four parts: the diagram, the input $x$ to $u$, positive
integers $p$ (position) and $t$ (time in the execution of $u(x)$):}
\hspace{1pc} \fbox {$\begin{array}{|c|c|c|c|c|c|c}\cline{1-2}\cline{5-5}
 p&t&\multicolumn{1}{c}{}&&A&\multicolumn{1}{c}{}&t{+}1\\
 \cline{1-2}\cline{4-6}\multicolumn{2}{|c|}{x}&&B_{-1}&B_0&B_{+1}&t\\
 \cline{1-2}\cline{4-6} \multicolumn{3}{c}{}& \multicolumn{1}{c}{p{-}1}&
 \multicolumn{1}{c}{p}& \multicolumn{1}{c}{p{+}1}\end{array}$}

The diagram shows the states $A$ of cell $p$ at time $t{+}1$ and $B_s,\,s{\in}
\{0,\pm1\}$ of cells $p{+}s$, at time $t$. $A,B$ include the pointer direction;
$B$ may be replaced by ``?". Some board configurations are illegal:
if (1) two of $B_s$ point away from each other, or (2) $A$ differs from
the result prescribed by the transition rules for $B_s$, or (3) $t=1$,
while $(B_s) \ne x_{p{+}s}$. (At $t=1$, $u(x)$ is just starting, so its tape
has the input $x$ at the left starting with the head in the initial state,
followed by blanks at the right.) Here are the {\bf Game Rules:}

The game starts in the configuration shown below.
 $L$ moves first, replacing the ?s with symbols that claim to reflect the state
of cells $p{+}s$ at step $t$ of $u(x)$. $S$ in its move chooses $s$, copies
$B_s$ into $A$, fills all $B$ with ?s, adds $s$ to $p$, and decrements $t$.

\noindent Start: \fbox{$\begin{array}{|c|c|c|c|c|c|}\cline{1-2}\cline{5-5}
 p=0&t=2^{\|x\|}&\multicolumn{1}{c}{}&&\leftarrow&\multicolumn{1}{c}{}\\
 \cline{1-2}\cline{4-6}\multicolumn{2}{|c|}{\text{input}\ x}&&?&?&?\\
 \cline{1-2}\cline{4-6}\end{array}$} \hspace{1pc}
 L puts: \fbox{$\begin{array}{|c|c|c|c}\cline{2-2}
 \multicolumn{1}{c|}{}&A&\multicolumn{1}{c}{}&t{+}1\\
 \cline{1-3}B_{-1}&B_0&B_{+1}&t\\\cline{1-3}\end{array}$} \hspace{1pc}
 S puts: \fbox{$\begin{array}{|c|c|c|c}\cline{2-2}
 \multicolumn{1}{c|}{}&B_s&\multicolumn{1}{c}{}&t\\
 \cline{1-3}?&?&?&t{-}1\\\cline{1-3}\end{array}$}

Note that $L$ may lie (i.e fill in ``?" distorting the actual computation
of $u(x)$), as long as he is consistent with the above ``local" rules.
All $S$ can do is to check the two consecutive board configurations.
He cannot refer to past moves or to actual computation
of $u(x)$ as an evidence of $L$'s violation.

\paragraph {Strategy:} If $u(x)$ does indeed halt within $2^{\|x\|}$ steps,
then the initial configuration is true to the computation of $u(x)$.
Then $L$ has an obvious (though hard to compute) winning strategy: just tell
truly (and thus always consistently) what actually happens in the computation.
$S$ will lose when $t{=}1$ and cannot decrease any more. If the initial
configuration is a lie, $S$ can force $L$ to lie all the way down to $t=1$. How?

If the upper box $A$ of a legal configuration is false then the lower boxes
$B_s$ cannot all be true, since the rules of $u$ determine $A$ uniquely from
them. If $S$ correctly points the false $B_s$ and brings it to the top on his
move, then $L$ is forced to keep on lying. At time $t{=}1$ the lie is exposed:
the configuration doesn't match the actual input string $x$, i.e. is illegal.

Solving this game amounts to deciding correctness of the initial configuration,
i.e. $u(x)$ halting in $2^{\|x\|}$ steps: impossible in time $o(2^{\|x\|})$.
This Halting Game is artificial, still has a BHP flavor,
though it does not refer to exponents. We now reduce it
to a nicer game (Linear Chess) to prove it exponentially hard, too.

\newpage\subsection {Reductions; Non-Deterministic and Alternating TM;
 Time and Space.}\label{gm-reduce}

To \trm {reduce} (see definition in sec.~\ref{rec})
Halting game to Linear Chess we introduce a few concepts.

A \trm {non-deterministic} Turing Machine (NTM) is a TM that sometimes offers a
(restricted) transition choice, made by a \trm {driver}, a function (of the TM
configuration) of unrestricted complexity. A deterministic (ordinary) TM $M$
accepts a string $x$ if $M(x){=}$yes; an NTM $M$ does if there exists a driver
$d$ s.t. $M_d(x){=}$yes. NTM represent single player games -- puzzles -- with a
simple transition rule, e.g., Rubik's Cube.\\ One can compute the winning
strategy in exponential time by exhaustive search of all $d$.

{\bf Home Work:} Prove all such games have P-time winning strategies,
or show some have not.\\ Will get you grade A for the course,
\$1,000,000 Award and a senior faculty rank at a school of your choice.

The \trm {alternating} TM (ATM) is a variation of the NTM driven by two
alternating drivers (players) $l,r$. A string is accepted if there is $l$ such
that for any $r: M_{l,r}(x){=}$ yes. Our games could be viewed as ATM returning
the result of the game in linear space but possibly exponential time, $M$
prompts $l$ and $r$ alternatingly to choose their moves (in several steps if
the move is specified by several bits) and computes the resulting position,
until a winner emerges. Accepted strings describe winning positions.

\vspace{-8pt}\paragraph {Linear Chess (LC)  Simulation of TM-Games.\label{dc1}}
The simulation first represents the Halting Game as an ATM computation
simulated by the Ikeno TM (\ref{ikeno}) (using the ``A/B" command for players'
input). The UTM is viewed as an array of 1-pointer cellular automata:
Weak cells as rightward, Shy leftward. Upon termination, the TM head
is set to move to the end of the tape, eliminating all loser pieces.

This is viewed as a game of \trm {1d-Chess (1dC)}, a variant of LC, where the
table, not the ``Gender Rule", determine the victorious piece, and not only
the vanquished piece is replaced, but also the winning piece may be ``promoted
to" another type of the same side. The types are states of Ikeno TM showing
Loyalty (pointer direction) $d{\in}\{W,S\}$, gender $g$ (= previous $d$), and
6/6/6/5 ranks (trit $t{\in}\{0,1,*\}$ with $'$ bit $p$).

{\bf Exercise:} Describe LC simulation of 1dC. {\bf Hint:} 
Each 1dC transition is simulated in several LC stages.
Let L,R be the two pieces active in 1dC. In odd stages L (in even stages R)
changes gender while turning pointer twice. The last stage turns pointer
only once and possibly changes gender. In the first stage L appends its rank
with R's $p$ bit. All other stages replace old 1dC rank with the new one.
R appends its old $t$ bit (only if $t\ne*$) to its new rank. Subsequent
stages drop both old bits, marking L instead if it is the new 1dC head.
Up to 4 more stages are used to exit any mismatch with 1dC new $d,g$ bits.

\paragraph {Space-Time Trade-off.}
\trm {Deterministic} linear space computations are games
where any position has at most one (and easily computable) move.
We know no general superlinear lower bound or subexponential upper bound
for time required to determine their outcome. This is a big open problem.

\noindent\parbox{296pt}{Recall that on a parallel machine: \trm {time} is the
number of steps until the last processor halts; \trm {space} is the amount of
memory used; \trm {volume} is the combined number of steps of all processors.
``\trm {Small}" will refer to values bounded by a polynomial of the input
length; ``\trm {large}" to exponential. Let us call computations \trm {narrow}
if {\em either} time {\em or} space are polynomial, and \trm {compact} if both
(and, thus, volume too) are. {\bf An open question:} Do all exponential volume
algorithms (e.g., one solving Linear Chess) allow an equivalent {\em narrow}
computation?} \hspace{1pc}\parbox{157pt} {\vector(1,0){130} space\\
 \mbox{\raisebox{75pt} {\vector(0,-1){70}}\hspace{-9pt}\raisebox{-4pt} {time}
 \fbox{\rule{0pt}{6pc}\parbox[b]{2pc} {large\\time,\\small\\space}}\hspace{3pt}
 \raisebox{43pt} {\parbox{106pt} {\fbox {\rule{0pt}{18pt}small time,
 large space}\vspace{2pc} narrow computations}}}}\vspace{6pt}

{\bf Alternatively, can every {\em narrow} computation be converted into
 a compact one?} This is equivalent to the existence of a P-time algorithm
for solving any \trm {fast} game, i.e. a game with a P-time transition rule
and a move counter limiting {\em the number of moves} to a polynomial. The
sec.~\ref{win} algorithm can be implemented in parallel P-time for such games.
Converse also holds, similarly to the Halting Game.

\cite{stm} solve compact games in P-space: With $M{\subset}\{0,1\}$ run
depth-first search on the tree of all games -- sequences of moves.
On exiting each node it is marked as the active player's win if some move leads
to a child so marked; else as his opponent's. Children's marks are then erased.
Conversely, compact games can simulate any P-space algorithms.
Player A declares the result of the space-$k$, time-$2^k$ computation.
If he lies, player B asks him to declare the memory state in the middle of
that time interval, and so by a k-step binary search catches A's lie on a
mismatch of states at two adjacent times. This has some flavor of trade-offs
such as saving time at the expense of space in dynamic programming.

Thus, fast games (i.e. compact alternating computations) correspond
to narrow deterministic computations; general games (i.e. narrow
alternating computations) correspond to large deterministic ones.

\newpage\part {Mysteries.} \label{myst}
We now enter Terra incognita by extending deterministic computations with tools
like random choices, non-deterministic guesses, etc., the power of which is
completely unknown. Yet many fascinating discoveries were made there in which
we will proceed to take a glimpse.

\section
{Nondeterminism; Inverting Functions; Reductions.} \label{np}

\subsection {An Example of a Narrow Computation:
 Inverting a Function.}\label {invert}

Consider a P-time function $F$. For convenience, assume $\|F(x)\|=\|x\|$,
(often $\|F(x)\|=\|x\|^{\Theta(1)}$ suffices). Inverting $F$ means finding,
for a given $y$, at least one $x\in F^{-1}(y)$, i.e. such that $F(x) = y$.

We may try all possible $x$ for $F(x)=y$. Assume $F$ runs in linear time on
a Pointer Machine. What is the cost of inverting $F$? The {\em space} used
is $\|x\|+\|y\|+ $space$_F(x)= O(\|x\|)$. But time is $O(\|x\|2^{\|x\|})$:
absolutely infeasible. No method is currently proven much better in the worst
case. And neither could we prove some inversion problems to require
{\em super-linear} time. This is the sad present state of Computer Science!

\vspace{-7pt}\paragraph {An Example: Factoring.} Let $F(x_1,x_2)=x_1x_2$ be
the product of integers. For simplicity, assume $x_1,x_2$ are primes.
A fast algorithm in sec.~\ref{prime} determines if an integer is prime. If not,
no factor is given, only its existence. To invert $F$ means to factor $F(x)$.
The density of $n$-bit primes is $\approx1/(n\ln2)$.
So, factoring by exhaustive search takes {\em exponential} time!
In fact, even the best known algorithms for this ancient problem run in
time about $2^{\sqrt{\|y\|}}$, despite centuries of efforts by most brilliant
people. The task is now commonly believed infeasible and the security
of many famous cryptographic schemes depends on this {\em unproven} faith.

\vspace{-7pt}\paragraph {One-Way Functions:} $F:x\to y$ are those easy to
compute $(x\mapsto y)$ and hard to invert $(y\mapsto x)$ for most $x$.
Even their existence is sort of a religious belief in Computer Theory.
It is unproven, though many functions {\em seem} to be one-way.
Some functions, however, are proven to be one-way, IFF one-way functions EXIST.
Many theories and applications are based on this hypothetical existence.

\subsubsection*{Search and NP Problems.} \label{search}

Let us compare the inversion problems with another type -- the search problems
specified by computable\\ in time $\|x\|^{O(1)}$ relations $P(x,w)$:
given $x$, find $w$ s.t. $P(x,w)$. There are two parts to a search problem:\\
 (a) decision problem: decide if $w$ (called \trm{witness}) exist, and (b) a
constructive problem: actually find $w$.

Any inversion problem is a search problem and any search problem can be
restated as an inversion problem. E.g., finding a Hamiltonian cycle $C$ in a
graph $G$, can be stated as inverting a $f(G,C)$, which outputs $G,0\ldots 0$
if $C$ is in fact a Hamiltonian cycle of $G$. Otherwise, $f(G,C) = 0\ldots 0$.

Similarly any search problem can be reduced to another one equivalent to
its decision version.\\ For instance, factoring $x$ reduces to bounded
factoring: given $x,b$ find $p,q$ such that $pq=x$, $p\le b$\\
(where decisions yield construction by binary search).

{\bf Exercise:} Generalize the two above examples to reduce any search
problem to an inverting problem and to a decision problem.

The \trm {language} of a problem is the set of all acceptable inputs.
For an inversion problem it is the range of $f$. For a search problem it is
the set of all $x$ s.t. $P(x,w)$ holds for some $w$. An \trm {NP language} is
the set of all inputs acceptable by a P-time \trm {non-deterministic} Turing
Machine (sec.~\ref{gm-reduce}). All three classes of languages -- search,
inversion and NP -- coincide (NP $\iff$ search is straightforward).

Interestingly, polynomial {\em space} bounded deterministic and
non-deterministic TMs have equivalent power. It is easy to modify TM to have
a unique accepting configuration. Any acceptable string will be accepted in
time $2^s$, where $s$ is the space bound. Then we need to check $A(x,w,s,k)$:
whether the TM can be driven from the configuration $x$ to $w$ in time $<2^k$
and space $s$. For this we need for every $z$, to check $A(x,z,s,k{-}1)$
and $A(z,w,s,k{-}1)$, which takes space $t_k\le t_{k{-}1} + \|z\|+O(1)$.
So, $t_k= O(sk)= O(s^2)$ \cite{Sv}.

Search problems are games with P-time transition rules and one move duration.
A great hierarchy of problems results from allowing more moves
and/or other complexity bounds for transition rules.

\newpage\subsection{Complexity of NP Problems.}\label{compl}

We discussed the (equivalent) inversion, search, and NP types of problems.
Nobody knows whether {\em all} such problems are solvable in P-time (i.e.
belong to P). This question (called P=?NP) is probably the most famous one
in Theoretical Computer Science. All such problems are solvable in exponential
time but it is unknown whether any better algorithms generally exist. For many
problems the task of finding an efficient algorithm may seem hopeless,
while similar or slightly modified problems have been solved. Examples:

 \begin{enumerate} \itemsep0pt
 \item Linear Programming: Given integer $n\times m$
  matrix $A$ and vector $b$, find a rational vector $x$ with $Ax<b$.
  Note, if $n$ and entries in $A$ have $\le k$-bits
  and $x$ exists then an $O(nk)$-bit $x$ exists, too.

Solution: The Dantzig's \trm {Simplex} algorithm finds $x$
quickly for many $A$.\\ Some $A$, however, take exponential time.
After long frustrating efforts, a worst case\\
P-time Ellipsoid Algorithm was finally found in \cite{YN}.

\item Primality test: Determine whether a given integer $p$ has a factor?

Solution: A bad (exponential time) way is to try
all $2^{\|p\|}$ possible integer factors of $p$.\\
More sophisticated algorithms, however, run fast (see section~\ref{prime}).

\item Graph Isomorphism Problem: Are two given graphs $G_1,G_2$, isomorphic?\\
 I.e., can the vertices of $G_1$ be re-numbered so that it becomes equal $G_2$?

Solution: Checking all $n!$ enumerations of vertices is impractical\\
 (for $n=100$, this exceeds the number of atoms in the known Universe).\\
 \cite{Luks} found an $O(n^d)$ steps algorithm where $d$ is the degree.
 This is a P-time for $d = O(1)$.

\item Independent Edges (Matching):\\ Find a given number of
 independent (i.e., not sharing nodes) edges in a given graph.

Solution: Max flow algorithm solves a bipartite graph case.\\ The general case
is solved with a more sophisticated algorithm by J. Edmonds. \end{enumerate}

Many other problems have been battled for decades or centuries
and no P-time solution has been found. Even modifications of
the previous four examples have no known answers:

 \begin{enumerate}\itemsep0pt
 \item Linear Programming: All known solutions produce rational $x$.\\
 No reasonable algorithm is known to find integer $x$.
 \item Factoring: Given an integer, find a factor.
 Can be done in about exponential time $n^{\sqrt{n}}$.\\
 Seems very hard: Centuries of quest for fast algorithm were unsuccessful.
 \item Sub-graph isomorphism: In a more general case of finding isomorphisms
 of a graph\\ to a part of another, no P-time solution has been found,
 even for $O(1)$-degree graphs.
 \item Independent Nodes: Find $k$ independent
 (i.e., not sharing edges) nodes in a given graph.\\
 No P-time solution is known. \end{enumerate}

{\bf Exercise:}
 Restate the above problems as inverting easily computable functions.

We learned the proofs that Linear Chess and some other games have exponential
complexity. None of the above or any other search/inversion/NP problem,
however, have been proven to require super-P-time.
When, therefore, do we stop looking for an efficient solution?

\paragraph {NP-Completeness} theory is an attempt to answer this question.\\
See results by S.Cook, R.Karp, L.Levin, and others surveyed in \cite{GJ,trakh}.

A P-time function $f$ reduces one NP-predicate $p_1(x)$ to $p_2(x)$ iff $p_1(x)
= p_2(f(x))$, for all $x$. $p_2$ is NP-complete if {\em all} NP problems can
be reduced to it. Thus, each NP-complete problem is at least as worst-case
hard as all other NP problems. This may be a good reason to give up on fast
algorithms for it. Any P-time algorithm for one NP-complete problem would yield
one for all other NP (or inversion, or search) problems. No such solution
has been discovered yet and this is left as a homework (10 years deadline).

Faced with an NP-complete problem we can sometimes restate it, find a similar
one which is easier (possibly with additional tools) but still gives the
information we really want. We will do this in Sec.~\ref{prime} for factoring.
Now we proceed with an example of NP-completeness.

\newpage\subsection {An NP-Complete Problem: Tiling.}\label{tile}

\parbox{32pc} {{\bf Tiling Problem.} Invert the function which, given a tiled
square, outputs its first row and the list of tiles used. A tile is one of
the $26^4$ possible squares containing a Latin letter at each corner. Two tiles
may be placed next to each other if the letters on the shared side match. 
(See an example at the right.) We now reduce to Tiling any search problem:
given $x$, find $w$ satisfying a P-time computable property $P(x,w)$.}
 \hspace{9pt} \parbox{6pc} {\mbox
 {\fbox{\parbox{2pc} {\makebox[2pc] {a\hfill x} \makebox[2pc] {m\hfill r}}}
 \fbox{\parbox{2pc} {\makebox[2pc] {x\hfill c} \makebox[2pc] {r\hfill z}}}}
 \par \vspace {3pt} \mbox
 {\fbox{\parbox{2pc} {\makebox[2pc] {m\hfill r} \makebox[2pc] {n\hfill s}}}
 \fbox{\parbox{2pc} {\makebox[2pc] {r\hfill z} \makebox[2pc] {s\hfill z}}}}}

\paragraph {Padding Argument.} First, we need to reduce it to some
``standard" NP problem. An obvious candidate is the problem ``Is there $w:
U(v,w)$ ?", where $U$ is the universal Turing Machine, simulating $P(x,w)$ for
$v = px$. But $U$ does not run in P-time, so we must restrict $U$ to $u$ which
stops within some P-time limit. How to make this fixed degree limit sufficient
to simulate any polynomial (even of higher degree) time $P$? Let the TM
$u(v,w)$ for $v{=}00\ldots01px$ simulate $\|v\|^2$ steps of $U(px,w){=}P(x,w)$.
If the \trm {padding} of $0$'s in $v$ is sufficiently long, $u$ will have
enough time to simulate $P$, even though $u$ runs in quadratic time, while
$P$'s time limit may be, say, cube (of a shorter ``un-padded" string).
So the NP problem $P(x,w)$ is reduced to $u(v,w)$ by mapping instances $x$
into $f(x)= 0\ldots01px=v$, with $\|v\|$ determined by the time limit for $P$.
Notice that program $p$ for $P$ is fixed. So, if some NP problem {\em cannot}
be solved in P-time then neither can be the problem $\exists? w: u(v,w)$.
Equivalently, if the latter {\em IS} solvable in P-time then so is {\em any}
search problem. We do not know which of these alternatives is true.
It remains to reduce the search problem $u$ to Tiling.

\paragraph {The Reduction.} We compute $u(v,w)$ (where $v = 00\ldots01px$)
by a TM represented as an array of 1-pointer cellular automata that runs
for $\|v\|^2$ steps and stops if $w$ does NOT solve the relation $P$.
Otherwise it enters an infinite loop. An instance $x$ has a solution
iff $u(v,w)$ runs forever for some $w$ and $v = 0\ldots01px$.

\noindent \parbox{297pt} {Here is the space-time diagram of computation of
$u(v,w)$. We set $n$ to $u$'s time (and space) $\|v\|^2$. Each row in this
table represents the configuration of $u$ in the respective moment of time.
 The solution $w$ is filled in at the second step below a special symbol "?".
If a table is filled in wrongly, i.e. doesn't reflect any actual computation,
then it must have four cells sharing a corner that couldn't possibly
appear in the computation of $u$ on any input.}
 \hspace{1pc} \parbox{13pc} {\vector(1,0){89}~space:~$n{=}\|v\|^2$\\\mbox
 {\raisebox{3pc}{\vector(0,-1){50}}\hspace{-10pt}\raisebox{-2pc}{time} \fbox
 {\parbox{11pc}{\makebox[11pc]{$v$ ?\ldots? \#\ldots\#\hfill(init.config.)}
\makebox[11pc]{\makebox[2pc]{$v$\hfill$w$}\hfill $T_1$\hfill\makebox[2pc]{}}
 \makebox[11pc]{\vdots\hfill\vdots\hfill\vdots} \makebox[11pc]{$T_n$}}}}}

\paragraph {Proof.} As the input $v$ and the guessed solution $w$ are
the same in both the right and the wrong tables, the first 2 rows agree.
The actual computation starts on the third row. Obviously, in the first
mismatching row a transition of some cell from the previous row is wrong.
This is visible from the state in both rows of this cell and the cell it points
to, resulting in an impossible combination of four cells sharing a corner.

\noindent\parbox{383pt} {\hspace*{1pc} For a given $x$, the existence of $w$
satisfying $P(x,w)$ is equivalent to the existence of a table with the
prescribed first row, no halting state, and permissible patterns of each four
adjacent squares (cells). Converting the table into the \trm{Tiling Problem}:\\
The cells in the table are separated by ``---" ; the tiles by ``...";
Cut each cell into 4 parts by a vertical and a horizontal lines
through its center and copy cell's content in each part.
Combine into a tile each four parts sharing a corner of 4 cells.
If these cells are permissible in the table, then so is the respective tile.}
 \hspace{2pc} $\raisebox{-18pt}{\dashbox{1}(40,40){\parbox{2.5pc}
 {\makebox[2.5pc]{u\hfill v}\\\hspace*{2.5pc}\\\makebox[2.5pc]{v\hfill x}}}}
\hspace{-57pt}\parbox{75pt} {\framebox[3pc]{\rule{0pt}{2pc}}
\framebox[3pc]{\rule{0pt}{2pc}}\vspace{3pt} \framebox[3pc]{\rule{0pt}{2pc}}
\framebox[3pc]{\rule{0pt}{2pc}}}$

\vspace{6pt} So, any P-time algorithm extending a given first row
to the whole table of matching tiles from a given set could be
used to solve any NP problem by converting it to Tiling as shown.

\paragraph {Exercise:}
Find a polynomial time algorithm for $n\times\log n$ Tiling Problem.

\newpage\section {Probability in Computing.}\label{prob}

\subsection {A Monte-Carlo Primality Tester.}\label{prime}

The factoring problem seems very hard. But to test if a number has factors
turns out to be much easier than to find them. It also helps if we supply
the computer with a coin-flipping device. See: \cite{rabin,miller,so-st}.
We now consider a Monte Carlo algorithm, i.e. one that with high
probability rejects any composite number, but never a prime.

\paragraph{Residue Arithmetic.} $p|x$ means $p$ divides $x$.
$x\equiv y\pmod p$ means $p|(x{-}y)$. $y=(x \bmod p)$ denotes the residue of
$x$ when divided by $p$, i.e. $x\equiv y\in[0,p{-}1]$. Residues can be added,
multiplied and subtracted with the result put back in the range $[0,p {-}1]$
via shifting by an appropriate multiple of $p$. E.g., $-x$ means $p{-}x$
for residues $\bmod\ p$. We use $\pm x$ to mean either $x$ or $-x$.

The Euclidean Algorithm finds $\gcd(x,y)$ -- the greatest (and divisible
by any other) common divisor of $x$ and $y$: $\gcd(x,0)=x$; $\gcd(x,y)=\gcd
(y,(x\bmod y))$, for $y>0$. By induction, $g{=}\gcd(x,y){=}A*x{-}B*y$,
where integers $A{=}(g/x\bmod y)$ and $B{=}(g/y\bmod x)$ are produced as
a byproduct of Euclid's Algorithm. This allows division $(\bmod\ p)$ by any $r$
\trm {coprime} with $p$, (i.e. $\gcd(r,p){=}1$), and operations $+,-,*,/$
obey all usual arithmetical laws. We will need to compute $(x^q\bmod p)$
in polynomial time. We cannot do $q{>}2^{\|q\|}$ multiplications. Instead
we compute all numbers $x_i=(x_{i{-}1}^2\bmod p)= (x^{2^i}\bmod p),i<\|q\|$.
Then we represent $q$ in binary, i.e. as a sum of
powers of $2$ and multiply $\bmod\ p$ the needed $x_i$'s.

\paragraph{Fermat Test.} The Little Fermat Theorem for every
prime $p$ and $x\in[1,p{-}1]$ says: $x^{(p{-}1)}\equiv1\pmod p$.\\
 Indeed, the sequence $(xi\bmod p)$ is a permutation of $1,\ldots,p{-}1$.
 So, $1{\equiv}(\prod_{i<p}(xi))/(p{-}1)!\equiv x^{p-1} (\bmod p)$.

 This test rejects typical composite $p$.
 Other composite $p$ can be actually factored by the following tests.

\paragraph {Square Root Test.} For each y and prime p,
 $x^2\equiv y\pmod p$ has at most one pair of solutions $\pm x$.

{\bf Proof.} Let $x,x'$ be two solutions: $y\equiv x^2\equiv x'^2\pmod
p$. Then $x^2-{x'}^2 = (x{-}x')(x{+}x')\equiv0\pmod p$.\\ So, $p$ divides
$(x{-}x') (x{+}x')$ and, if prime, must divide either $(x{-}x')$ or
$(x{+}x')$.\\ (Thus either $(x\equiv x')$ or $(x\equiv-x')$.) Otherwise
 $p$ is composite, and $\gcd(p,x{+}x')$ {\em actually gives} its factor.

\paragraph {Miller-Rabin Test} $T(x,p)$ completes the Fermat Test:
it factors a composite $p$, given $d$ that \\\trm {kills} Z$_p^*$
(i.e. $x^d\equiv\gcd(x,p)^d\pmod p$ for all $x$) and a random choice of $x$.
For prime $p$, $d=p{-}1$.

Let $d{=}2^k q$, with odd $q$. $T$ sets $x_0=(x^q\bmod p)$,
$x_i=(x_{i{-}1}^2\bmod p)=(x^{2^i q}\bmod p)$, $i\le k$. If $x_k{\ne}1$ then
$\gcd(x,p){\ne}1$ factors $p$ (if $d$ killed $x$, else Fermat test rejects
$p{=}d{+}1$). If $x_0{=}1$, or one of $x_i$ is $-1$, $T$ gives up for this $x$.
 Otherwise $x_i{\notin}\{\pm1\}$ for some $i{<}k$, while
 $(x_i^2\bmod p){=}x_{i+1}{=}1$, and the Square Root Test factors $p$.

First, for each odd composite $p$, we show that $T$ succeeds with {\em some}
$x$, coprime with $p$. If $p{}=a^j,j{>}1$, then $x{=}(1{+}p/a)$ works for Fermat
Test: $(1{+}p/a)^{p{-}1}{=}1{+}(p/a)(p{-}1){+}(p/a)^2(p{-}1)(p{-}2)/2{+}\ldots
\equiv 1{-}p/a{\not\equiv}1\pmod p$, since $p|(p/a)^2$. Otherwise $p{=}ab,
\gcd(a,b){=}1{<}a{<}b$. Take the {\bf greatest} $i$ such that $x_i{\not\equiv}1$
for some $x$ coprime with $p$. It exists: $(-1)^q\equiv-1$ for odd $q$.
So, $(x_i)^2\equiv 1\not\equiv x_i\pmod p$. (Or $i{=}k$, so Fermat test works.)
Then $x'{=}1{+} b(1/b\bmod a)(x{-}1)\equiv1{\equiv}x'_i\pmod b$, while
$x'_i{\equiv}x_i\pmod a$. So, either $x_i$ or $x'_i$ is $\not\equiv\pm1\pmod p$.

Now, $T(y,p)$ succeeds with {\em most} $y_i$, as it does with $x_i$ (or $x'_i$):
the function $y\mapsto xy$ is 1-1 and $T$ cannot fail with both $y$ and $xy$.
This test can be repeated for many randomly chosen $y$. Each time $T$ fails, we
are twice more sure that $p$ is prime. The probability of $300$ failures on a
composite $p$ is $<2^{-300}$, its inverse exceeds the number of atoms in the
known Universe.

\newpage\subsection
 {Randomized Algorithms and Random Inputs.}\label{average}

\trm {Las-Vegas} algorithms, unlike Monte-Carlo, never give wrong answers.
Unlucky coin-flips just make them run longer than expected. Quick-Sort is a
simple example. It is about as fast as deterministic sorters, but is popular
due to its simplicity. It sorts an array $a[1\ldots n]$ of $n>2$ numbers by
choosing in it a random \trm {pivot}, splitting the remaining array in two
by comparing with the pivot, and calling itself recursively on each half.

For easy reference, rename the array entries with their positions $1,\ldots,n$
in the {\em sorted output} (no effect on the algorithm). Denote $t(i)$ the
(random) time $i$ is chosen as a pivot. Then $i$ will ever be compared with $j$
iff either $t(i)$ or $t(j)$ is the smallest among $t(i),\ldots,t(j)$.
This has $2$ out of $|j{-}i|+1$ chances. So, the expected number of comparisons
is $\sum_{i,j>i} 2/(1{+}j{-}i)= -4n+ (n{+}1)\sum_{k=1}^n 2/k= 2n(\ln n-O(1))$.
 Note, that the expectation of the sum of variables is
 the sum of their expectations (not true, say, for product).

The above Monte-Carlo and Las-Vegas algorithms require choosing strings {\em at
random} with uniform distribution. We mentally picture that as flipping a coin.
(Computers use \trm {pseudo-random generators} rather than coins in hope,
rarely supported by proofs, that their outputs have all the statistical
properties of truly random coin flips needed for the analysis of the algorithm.)

\paragraph {Random Inputs} to Deterministic Algorithms are analyzed similarly
to algorithms that flip coins themselves and the two should not be confused.
Consider an example: Someone is interested in knowing whether or not certain
graphs contain Hamiltonian Cycles. He offers graphs and pays \$100 if we show
either that the graph {\em has} or that it {\em has not} Hamiltonian Cycles.
Hamiltonian Cycle problem is NP-Complete, so it should be very hard for {\em 
some}, but not necessarily for {\em most} graphs. In fact, if our patron chooses
the graphs uniformly, a fast algorithm can earn us the \$100 {\em most of the
time}! Let all graphs have $n$ nodes and, say, $d<\ln n/2$ mean degree and be
equally likely. Then we can use the following (deterministic) algorithm:\\
Output ``{\bf No} Hamiltonian Cycles" and collect the \$100, if the graph has
an isolated node. Otherwise, pass on that graph and the money. Now, how often
do we get our \$100. The probability that a given node $A$ of the graph is
isolated is $(1-1/n)^{dn}>(1-O(1/n))/\sqrt n$. Thus, the probability that
{\em none} of $n$ nodes is isolated (and we lose our \$100) is $O((1-1/\sqrt
n)^n)= O(1)/e^{\sqrt n}$ and vanishes fast. Similar calculations can be made
whenever $r = \lim (d/\ln n)<1$. If $r>1$, other fast algorithms can actually
find a Hamiltonian Cycle.\\ See: \cite{jnsn,karp-pr,gu}. See also \cite{vl}
for a proof that another graph problem is NP-complete even on average.
How do this HC algorithm and the above primality test differ?

\begin{itemize}\item The primality algorithm works for {\em all} instances.
It tosses the coin itself and can repeat it for a more reliable answer.
The HC algorithms only work for {\em most} instances
(with isolated nodes or generic HC). \item In the HC algorithms, we must
trust the customer to follow the presumed random procedure.\\ If he cheats
and produces rare graphs often, the analysis breaks down.\end{itemize}

\paragraph {Symmetry Breaking.} Randomness comes into Computer Science
in many other ways besides those we considered.
Here is a simple example: avoiding conflicts for shared resources.

{\bf Dining Philosophers.} They sit at a circular table.
Between each pair is either a knife or a fork, alternating. The problem is,
neighboring diners must share the utensils, cannot eat at the same time. How
can the philosophers complete the dinner given that all of them must act in the
same way without any central organizer? Trying to grab the knives and forks at
once may turn them into fighting philosophers. Instead they could each flip a
coin, and sit still if it comes up heads, otherwise try to grab the utensils.\\
If two diners try to grab the same utensil, neither succeeds.
If they repeat this procedure enough times,\\ most likely each philosopher
will eventually get both a knife and a fork without interference.

We have no time to actually analyze this and many other scenaria,
where randomness is crucial.\\
Instead we will take a look into the concept of Randomness itself.

\newpage\subsection {Arithmetization:
 One-Player Games with Randomized Transition.}\label{ip}

The results of section~\ref{games} can be extended to \trm {Arthur-Merlin}
games which have one player -- Merlin -- but a randomized transition function,
effectively using a dummy second player -- Arthur -- whose moves are just
coin-flips. We will reduce generic games to games in which any Merlin's
strategy in any losing position has exponentially small chance to win.

The trick achieving this, called \trm {arithmetization}, expresses
the boolean functions as low degree polynomials, and applies them
to $\Z_p$-tokens (let us call them \trm {bytes}) instead of bits.
It was proposed in Noam Nisan's article widely distributed over email in the
Fall of 1989 and quickly used in a flood of follow-ups for proving relations
between various complexity classes. We follow \cite{shamir,fl}.

Let $g$ be the (ATM-complete) game of 1d-Chess (\ref{dc1}), $r(m,x)$ with
$x{=}x_1\ldots x_s$, $m,x_i{\in}\{0,1\}$ be its transition rule. Configurations
include $x$ and a remaining moves counter $c\le2^s$. They are terminal if
$c{=}0$, winning to the player $x_1$. Intermediate configurations $(m,x,y)$
have $y$ claimed as a prefix of $r(m,x)$.

Let $t(m,x,y)$ be $1$ if $y{=}r(m,x)$, else $t{=}0$. 1d-Chess is simple, so $t$
can be expressed as a product of $s$ multilinear $O(1)$-sized terms, any
variable shared by at most two terms. Thus $t$ is a polynomial, quadratic in
each $m,x_i,y_i$. Let $V_c(x)$ be $1$ if the active player has a strategy to
win in the $c$ moves left, i.e. $V_0(x)\edf x_1$, $V_{c+1}(x)\edf$
$1{-}V_c(0,x,\{\})V_c(1,x,\{\})= 1{-}V_c(r(0,x))V_c(r(1,x))$, where
$V_c(m,x,y)\edf V_c(y)t(m,x,y)$ for $y=y_1\ldots y_s$ or
$V_c(m,x,y)\edf V_c(m,x,y{\circ}0){+}V_c(m,x,y{\circ}1)$ for shorter $y$.
($\circ$ stands for concatenation.)

$G$ will allow Merlin to prove $x$ is winning i.e., $V_c(x)=1$. Configurations
$X=(m,x,y,v)$ of $G$ replace bits with $\Z_p$ bytes and add $v{\in}\Z_p$
reflecting Merlin's claimed $V$. The polynomial $V_c(m,x,y)$ is quadratic in
each $x_i,m$, as $t(m,x,y)$ is. Then $V_c(y)$ has degree 4 in $y_i$ and
$V_c(m,x,y)$ has degree 6 in $y_i$.

Merlin starts with choosing a $2s$-bit prime $p$. Then at each step with
$X=(m,x,y,v)$ Merlin gives an $O(1)$-degree polynomial $P$ with
$P(1)P(0)=1{-}v$ for $s$-byte $y$ or $P(0){+}P(1)=v$ for shorter $y$. Arthur
then selects a random $r{\in}\Z_p$ and $X$ becomes $(r,y,\{\},P(r))$ for
$s$-byte $y$ or $(m,x,y{\circ}r,P(r))$ for shorter $y$.

If the original $v$ is correct, then Merlin's obvious winning strategy
is to always provide the correct polynomial. If the original $v$ is wrong
then either $P(1)$ or $P(0)$ must be wrong, too, so they will agree only
with a wrong $P$. A wrong $P$ can agree with the correct one only
on few (bounded by degree) points. Thus it will give a correct value only to
exponentially small fraction of random $r$. Thus the wrong $v$ will propagate
throughput the game until it becomes obvious in the terminal configuration.

\vspace{2pc}
This reduction of Section~\ref{games} games yields a hierarchy of Arthur-Merlin
games powers, i.e. the type of computations that have reductions to $V_c(x)$ of
such games and back. The one-player games with randomized transition rule $r$
running in space linear in the size of initial configuration are equivalent to
exponential time deterministic computations. If instead the running time $T$ of
$r$ combined for all steps is limited by a polynomial, then the games are
equivalent to polynomial space deterministic computations.

An interesting twist comes in one move games with polylog $T$, too tiny
to examine the initial configuration $x$ and the Merlin's move $m$.
 But not only this obstacle is removed but the equivalence to NP is achieved
with a little care. Namely, $x$ is set in an error-correcting code, and $r$ is
given $O(\log\|x\|)$ coin-flips and random access to the digits of $x,m$.
Then the membership proof $m$ is reliably verified by the randomized $r$.\\
 See \cite{holo} for details and references.

\newpage\section {Randomness}\label{rand}

\subsection {Randomness and Complexity.}\label{kolm}

Intuitively, a random sequence is one that has the same properties as
a sequence of coin flips. But this definition leaves the question,
what {\em are} these properties? Kolmogorov resolved these problems with
a new definition of random sequences: those with no description noticeably
shorter than their full length. See survey and history in \cite{ku87,vitan}.

\paragraph {Kolmogorov Complexity} $K_A(x|y)$ of the string $x$ given $y$ is
the length of the shortest program $p$ which lets algorithm $A$ transform $y$
into $x$: $\min\{(\|p\|):A(p,y)=x\}$. There exists a Universal Algorithm $U$
such that, $K_U(x)\le K_A(x)+O(1)$, for every algorithm $A$.
This constant $O(1)$ is bounded by the length of the program $U$ needs to
simulate $A$. We abbreviate $K_U(x|y)$ as $K(x|y)$, or $K(x)$ for empty $y$.

An example: For $A:x\mapsto x$, $K_A(x)=\|x\|$,
 so $K(x)<K_A(x)+O(1)<\|x\|+O(1)$.

Can we compute $K(x)$ by trying all programs $p,\|p\|{<}\|x\|{+}O(1)$
to find the shortest one generating $x$? This does not work because
some programs diverge, and the halting problem is unsolvable.\\ In fact,
no algorithm can compute $K$ or even any its lower bounds except $O(1)$.

Consider the Berry Paradox expressed in the phrase:
 ``The smallest integer which cannot\\ be uniquely and clearly
 defined by an English phrase of less than two hundred characters."\\
 There are $< 128^{200}$ English phrases of $< 200$ characters.
 So there must be integers not expressible\\ by such phrases and the
 smallest one among them. But isn't it described by the above phrase?

A similar argument proves that $K$ is not computable. Suppose an algorithm
$L(x)\ne O(1)$ computes a lower bound for $K(x)$. We can use it to compute
$f(n)$ that finds $x$ with $n< L(x)\le K(x)$, but $K(x)< K_f(x)+ O(1)$
and $K_f(f(n))\le \|n\|$, so $n< K(f(n))< \|n\|+O(1)= \log O(n)\ll n$: a
contradiction.\\ So, $K$ and its non-constant lower bounds are not computable.

An important application of Kolmogorov Complexity measures
the Mutual Information:\\ $I(x:y)= K(x)+ K(y)- K(x,y)$.
It has many uses which we cannot consider here.

\subsubsection*{Deficiency of Randomness.}

Some upper bounds of $K(x)$ are close in some important cases.
One such case is of $x$ generated at random. Define its \trm {rarity}
for uniform on $\{0,1\}^n$ distribution as $d(x)=n-K(x|n)\ge-O(1)$.

What is the probability of $d(x)>i$, for uniformly random $n$-bit $x$ ?
There are $2^n$ strings $x$ of length $n$.\\ If $d(x)>i$, then $K(x|n)< n-i$.
There are $<2^{n-i}$ programs of such length, generating $<2^{n-i}$ strings.\\
So, the probability of such strings is $<2^{n-i}/2^n= 2^{-i}$ (regardless of
$n$)! Even for $n= 1,000,000$,\\ the probability of $d(x)>300$ is absolutely
negligible (provided $x$ was indeed generated by fair coin flips).

Small rarity implies all other enumerable properties of random strings. Indeed,
let such property ``$x{\not\in}P$" have a negligible probability and $S_n$
be the number of $n$-bit strings violating $P$, so $s_n=\log(S_n)$ is small.\\
To generate $x$, we need only the algorithm enumerating $S_n$ and the $s_n$-bit
position of $x$ in that enumeration. Then the rarity $d(x)> n-(s_n{+}O(1))$
is large. Each $x$ violating $P$ will thus also violate the ``small rarity"
requirement. In particular, the small rarity implies unpredictability of bits
of random strings: A short algorithm with high prediction rate would assure
large $d(x)$. However, the randomness can only be refuted, cannot be confirmed:
we saw, $K$ and its lower bounds are not computable.

\paragraph {Rectification of Distributions.} We rarely have a source of
randomness with precisely known distribution. But there are very efficient ways
to convert ``roughly" random sources into perfect ones. Assume, we have such a
sequence with weird unknown distribution. We only know that its long enough
($m$ bits) segments have min-entropy $>k+i$, i.e. probability $<1/2^{k+i}$,
given all previous bits. (Without such $m$ we would not know a segment needed
to extract even one not fully predictable bit.) No relation is required between
$n,m,i,k$, but useful are small $m,i,k$ and huge $n=o(2^k/i)$. We can fold $X$
into an $n\times m$ matrix. We also need a small $m\times i$ matrix $Z$,
independent of $X$ and {\bf really} uniformly random (or random Toeplitz, i.e.
with restriction $Z_{a+1,b+1}=Z_{a,b}$). Then the $n\times i$ product $XZ$ has
uniform with accuracy $O(\sqrt{n i/2^k})$ distribution. This follows from
\cite{gl}, which uses earlier ideas of U. and V. Vazirani.

\newpage\subsection {Pseudo-randomness.} \label{pseudor}

The above definition of randomness is very robust, if not practical. True
random generators are rarely used in computing. The problem is {\em not} that
making a true random generator is impossible: we just saw efficient ways to
perfect the distributions of biased random sources. The reason lies in many
extra benefits provided by pseudorandom generators. E.g., when experimenting
with, debugging, or using a program one often needs to repeat the exact same
sequence. With a truly random generator, one actually has to record all its
outcomes: long and costly. The alternative is to generate \trm {pseudo-random}
strings from a short seed. Such methods were justified in \cite{bm,yao}:

First, take any one-way permutation $F_n(x)$ (see sec.~\ref{crypt}) with a
\trm {hard-core} bit (see below) $B_p(x)$ which is easy to compute from $x,p$,
but infeasible to guess from $p,n,F_n(x)$ with any noticeable correlation.\\
 Then take a random \trm {seed} of three $k$-bit parts $x_0,p,n$ and Repeat:
($S_i{\gets}B_p(x_i)$; $x_{i+1}{\gets}F_n(x_i)$; $i{\gets}i{+}1$).

We will see how distinguishing outputs $S$ of this generator
from strings of coin flips would imply the ability to invert $F$.
This is infeasible if $F$ is one-way. But if P=NP (a famous open problem),
no one-way $F$, and no pseudorandom generators could exist.

By Kolmogorov's standards, pseudo-random strings are not random: let $G$ be the
generator; $s$ be the seed, $G(s) = S$, and $\|S\|\gg k=\|s\|$. Then $K(S)\le
O(1)+k\ll\|S\|$, thus violating Kolmogorov's definition.\\ We can distinguish
between truly random and pseudo-random strings by simply trying all short
seeds. However this takes time exponential in the seed length. Realistically,
pseudo-random strings will be as good as a truly random ones if they can't
be distinguished in feasible time. Such generators we call \trm {perfect}.

\paragraph {Theorem:} \cite{yao} Let $G(s)=S\in\{0,1\}^n$ run in time $t_G$.
 Let a probabilistic algorithm $A$ in expected (over internal coin flips)
 time $t_A$ accept $G(s)$ and truly random strings with different by $d$
probabilities. Then, for random $i$, one can use $A$ to guess $S_i$
from $S_{i+1},S_{i+2}, \ldots$ in time $t_A+t_G$ with correlation $d/O(n)$.

\paragraph {Proof.} Let $p_i$ be the probability that $A$ accepts $S=G(s)$
modified by replacing its first $i$ digits\\ with truly random bits.
Then $p_0$ is the probability of accepting $G(s)$ and must differ by $d$ from\\
the probability $p_n$ of accepting random string. Then $p_{i-1}-p_i = d/n$, for
randomly chosen $i$.\\ Let $P_0$ and $P_1$ be the probabilities of accepting
$r0x$ and $r1x$ for $x=S_{i+1},S_{i+2},\ldots$, and random $(i{-}1)$-bit $r$.\\
Then $(P_1{+}P_0)/2$ averages to $p_i$, while $P_{S_i}=P_0{+}(P_1{-}P_0)S_i$
averages to $p_{i-1}$ and\\ $(P_1{-}P_0) (S_i{-}1/2)$ to $p_{i-1}{-}p_i=d/n$.
So, $P_1{-}P_0$ has the stated correlation with $S_i.\qed$

If the above generator was not perfect, one could guess $S_i$ from the sequence
$S_{i+1},S_{i+2},\ldots$\\ with a polynomial (in $1/\|s\|$) correlation.
 But, $S_{i+1}, S_{i+2}\ldots$ can be produced from $p,n,x_{i+1}$.\\
 So, one could guess $B_p(x_i)$ from $p,n,F(x_i)$ with correlation $d/n$,
 which cannot be done for hard-core $B$.

\paragraph {Hard Core.} The key to constructing a pseudorandom generator
is finding a hard core for a one-way $F$. The following $B$ is hard-core
for any one-way $F$, e.g., for Rabin's OWF in sec.~\ref{crypt}.\\
\cite{Knuth} has more details and references.

Let $B_p(x)=(x\cdot p)= (\sum_ix_ip_i\bmod2)$. \cite{gl} converts
any method $g$ of guessing $B_p(x)$ from $p,n,F(x)$ with correlation
$\varepsilon$ into an algorithm of finding $x$, i.e. inverting $F$
(slower $\varepsilon^2$ times than $g$).

\paragraph {Proof.} (Simplified with some ideas of Charles Rackoff.)
Take $k=\|x\|=\|y\|$, $j=\log(2k/\varepsilon^2)$, $v_i= 0^i10^{k-i}$.
Let $B_p(x) =(x\cdot p)$ and $b(x,p)=(-1)^{B_p(x)}$.
Assume, for $y=F_n(x)$, $g(y,p,w)\in\{\pm 1\}$ guesses $B_p(x)$
with correlation $\sum_p2^{-\|p\|}b(x,p) g_p >\varepsilon$, where $g_p$
abbreviates $g(y,p,w)$, since $w,y$ are fixed throughout the proof.

$(-1)^{(x\cdot p)}g_p$ averaged over ${>}2k/\varepsilon^2$ random pairwise
independent $p$ deviates from its mean (over all $p$) by ${<}\varepsilon$
(and so is ${>}0$) with probability $>1-1/2k$. The same for
$(-1)^{(x\cdot[p+v_i])} g_{p+v_i}= (-1)^{(x\cdot p)} g_{p+v_i} (-1)^{x_i}$.

Take a random $k\times j$ binary matrix $P$. The vectors $Pr$,
$r{\in}\{0,1\}^j\setminus\{0^j\}$ are pairwise independent. So,
for a fraction $\ge1-1/2k$ of $P$, sign$(\sum_r(-1)^{xPr}g_{Pr+v_i})=(-1)^{x_i}$.
We could thus find $x_i$ for all $i$ with probability $>1/2$
if we knew $z=xP$. But $z$ is short: we can try all its
$2^j$ possible values and check $y=F_n(x)$ for each !

So the inverter, for a random $P$ and all $i,r$, computes $G_i(r)=g_{Pr+v_i}$.
It uses Fast Fourier on $G_i$ to compute $h_i(z)=\sum_rb(z,r)G_i(r)$. The sign
of $h_i(z)$ is the $i$-th bit for the $z$-th member of output list. $\qed$

\newpage\subsection {Cryptography.} \label{crypt}

\paragraph {Rabin's One-way Function.} Pick random prime numbers $p,q,\|p\|=
\|q\|$ with two last bits ${=}1$, i.e. with odd $(p{-}1)(q{-}1)/4$. Then
$n=pq$ is called a Blum number. Its length should make factoring infeasible.

 Let $Q_n=(Z^*_n)^2$ be the set of squares,
 i.e. \trm {quadratic residues} (all residues are assumed $\pmod n$).

\paragraph {Lemma.} Let $n=pq$ be a Blum number, $F: x\mapsto x^2\in Q_n$.
Then (1) $F$ is a permutation on $Q_n$\\ and (2)
The ability to invert $F$ on random $x$ is equivalent to that of factoring $n$.

\vspace{-8pt}\paragraph {Proof.} (1) $t{=}(p{-}1)(q{-}1)/4$ is odd, so
$u{=}(t{+}1)/2$ is an integer. Let $x{=}F(z)$. Both $p{-}1$ and $q{-}1$
divide~$2t$. So, by Fermat's little theorem, both $p$, $q$ (and, thus $n$)
divide $x^t{-}1\equiv z^{2t}{-}1$. Then $F(x)^u\equiv x^{2u}=xx^t\equiv x$.

(2) The above $y^u$ inverts $F$. Conversely, let $F(A(y))=y$ for a fraction
$\varepsilon$ of $y\in Q_n$.\\ Each $y\in Q_n$ has $x,x'{\ne}\pm x$ with
$F(x){=}F(x'){=}y$, both with equal chance to be chosen at random.\\
If $F(x)$ generates $y$ while $A(y)=x'$ the Square Root Test
(\ref{prime}) has both $x,x'$ for factoring $n.\qed$

Such one-way permutations, called ``trap-door", have many applications;
we look at cryptography below.

Picking random primes is easy: they have density $1/O(\|p\|)$.
Indeed, one can see that $\binom{2n}n$ is divisible by every prime
$p{\in}[n,2n]$ but by no prime $p{\in}[\frac23n,n]$ or prime power $p^i{>}2n$.
So, $(\log\binom{2n}n)/ \log n=2n/\log n-O(1)$ is an upper bound on
the number of primes in $[n,2n]$ and a lower bound on that in $[1,2n]$
(and in $[3n,6n]$ as a simple calculation shows).
And fast VLSI exist to multiply long numbers and check primality.

\paragraph {Public Key Encryption.}

A perfect way to encrypt a message $m$ is to add it $\bmod2$ bit by bit to a
random string $S$ of the same length $k$. The resulting encryption $m \oplus S$
has the same uniform probability distribution, no matter what $m$ is. So it is
useless for the adversary who wants to learn something about $m$, without
knowing $S$. A disadvantage is that the communicating parties must share a
secret $S$ as large as all messages to be exchanged, combined. \trm {Public
Key} Cryptosystems use two keys. One key is needed to encrypt the messages and
may be completely disclosed to the public. The \trm {decryption} key must still
be kept secret, but need not be sent to the encrypting party. The same keys may
be used repeatedly for many messages.

Such cryptosystem can be obtained \cite{b-gw} by replacing the above random $S$
by pseudorandom $S_i= (s_i\cdot x)$; $s_{i+1} =(s_i^2\ \bmod n)$. Here a Blum
number $n=pq$ is chosen by the Decryptor and is public, but $p,q$ are kept
secret. The Encryptor chooses $x\in Z_2^{\|n\|},s_0\in Z_n$ at random and sends
$x,s_k, m{\oplus} S$. Assuming factoring is intractable for the adversary, $S$
should be indistinguishable from random strings (even with known $x,s_k$).
Then this scheme is as secure as if $S$ were random. The Decryptor
knows $p,q$ and can compute $u,t$ (see above) and $v=(u^{k-1}\bmod t)$.
So, he can find $s_1=(s_k^v\bmod n)$, and then $S$ and $m$.

Another use of the intractability of factoring is digital signatures
\cite{rsa,bb-sg}. Strings $x$ can be released as authorizations
of $y=(x^2\bmod n)$. Verifying $x$, is easy but the ability of
forging it for generic $y$ is equivalent to that of factoring $n$.

\vfill\subsubsection* {Go On!}

You noticed that most of our burning questions are still open. Take them on!

Start with reading recent results (FOCS/STOC is a good source).
See where you can improve them.\\ Start writing, first notes just for
your friends, then the real papers. Here is a little writing advice:

A well written paper has clear components: skeleton, muscles, etc.\\
The skeleton is an acyclic digraph of basic definitions and statements,
with cross-references.\\ The meat consists of proofs (muscles) each
{\em separately} verifiable by competent graduate students having to read
no other parts but statements and definitions cited. Intuitive comments,
examples and other comfort items are fat and skin: a lack or excess will
not make the paper pretty. Proper scholarly references constitute clothing,
no paper should ever appear in public without! Trains of thought
which led to the discovery are blood and guts: keep them hidden.
Metaphors for other vital parts, like open problems, I skip out of modesty.

\vfill\paragraph {Writing Contributions.} {\small
 Section~\ref{models} was originally prepared by Elena Temin,
 Yong Gao and Imre Kifor (BU), others by Berkeley students:
 \ref{compress} by Mark Sullivan,
 \ref{win} by Eric Herrmann and Elena Eliashberg,
 \ref{halt-gm} by Wayne Fenton and Peter Van Roy,
 \ref{gm-reduce} by Carl Ludewig, Sean Flynn, and Francois Dumas,
 \ref{invert} by Jeff Makaiwi, Brian Jones and Carl Ludewig,
 \ref{compl} by David Leech and Peter Van Roy,
 \ref{tile} by Johnny and Siu-Ling Chan, \ref{average} by Deborah Kordon,
 \ref{kolm} by Carl Ludewig, \ref{pseudor} by Sean Flynn,
 Francois Dumas, Eric Herrmann, \ref{crypt} by Brian Jones.}

\newpage\section {References}
 \renewcommand\refname{}

 \end{document}